\documentclass[prd,twocolumn,showpacs]{revtex4-1}
\usepackage{epsfig,amsmath,amssymb,mathtools,amscd}
\usepackage[linkcolor=red]{hyperref}
\usepackage{enumerate}
\newtheorem{lemma}{Lemma}

\newtheorem{proposition}{Proposition}

\mathchardef\minus="002D

\def\<{\langle}
\def\>{\rangle}
 \def\ket#1{| #1 \rangle}
\def\bra#1{\langle #1 |}
\def\ketbra#1#2{| #1 \rangle \langle#2 |}
\def\braket#1#2{\langle #1 | #2 \rangle}

\def\bvec#1{\boldsymbol{\mathrm #1}}

\def\Hilb{\mathcal H}

\newcommand\Item[1][]{%
  \ifx\relax#1\relax  \item \else \item[#1] \fi
  \abovedisplayskip=0pt\abovedisplayshortskip=0pt~\vspace*{-\baselineskip}}
\newcommand*\dd{\mathop{}\!\mathrm{d}}

\begin{document}
\title{The Thirring quantum cellular automaton}

\author{Alessandro 
  \surname{Bisio}} \email[]{alessandro.bisio@unipv.it}
\affiliation{Dipartimento di Fisica dell'Universit\`a di Pavia, via
  Bassi 6, 27100 Pavia} \affiliation{Istituto Nazionale di Fisica
  Nucleare, Gruppo IV, via Bassi 6, 27100 Pavia} 
\author{Giacomo Mauro
  \surname{D'Ariano}} \email[]{dariano@unipv.it}
\affiliation{Dipartimento di Fisica dell'Universit\`a di Pavia, via
  Bassi 6, 27100 Pavia} \affiliation{Istituto Nazionale di Fisica
  Nucleare, Gruppo IV, via Bassi 6, 27100 Pavia} 
\author{Paolo
  \surname{Perinotti}} \email[]{paolo.perinotti@unipv.it}
\affiliation{Dipartimento di Fisica dell'Universit\`a di Pavia, via
  Bassi 6, 27100 Pavia} \affiliation{Istituto Nazionale di Fisica
  Nucleare, Gruppo IV, via Bassi 6, 27100 Pavia} 
\author{Alessandro
  \surname{Tosini}}\email[]{alessandro.tosini@unipv.it}
\affiliation{Dipartimento di Fisica dell'Universit\`a di Pavia, via
  Bassi 6, 27100 Pavia} \affiliation{Istituto Nazionale di Fisica
  Nucleare, Gruppo IV, via Bassi 6, 27100 Pavia}

\begin{abstract} 
  We analytically diagonalize a discrete-time on-site interacting fermionic cellular automaton in the two-particle sector. Important features of the solutions sensibly differ from those of analogous Hamiltonian models. In particular, we found a wider variety of scattering processes, we have bound states for every value of the total momentum, and there exist bound states also in the free case, where the coupling constant is null.
\end{abstract} \pacs{71.10.Fd, 03.67.Ac}
\maketitle

Quantum cellular automata and quantum walks constitute an increasingly attractive arena for research in many body systems \cite{PhysRevLett.100.130501,PhysRevE.57.54,1742-5468-2017-8-083105}, quantum computation \cite{PhysRevLett.102.180501,PhysRevA.81.042330,PhysRevLett.104.050502,PhysRevLett.108.010502}, and foundations of quantum field theory \cite{bialynicki1994weyl,meyer1996quantum,bisio2013dirac,Arrighi2016,PhysRevA.90.062106}.

The notion of quantum cellular automaton introduced by Feynman \cite{feynman1982simulating} as a universal quantum simulator, was mathematically formalized in Ref.~\cite{schumacher2004reversible,gross2012index}. In the case of non-interacting theories the evolution of field operators is linear, and its simulation through quantum cellular automata reduces to simulation of a single particle through a quantum walk \cite{ambainis2001one,PhysRevA.48.1687,grossing1988quantum}. The interacting case is largely unexplored, and was mainly approached by extending the quantum walk formalism, introducing decoherence \cite{PhysRevA.81.062129}, or a classical external field \cite{bisio2013dirac,PhysRevA.93.052301,PhysRevA.94.012335,PhysRevE.55.5261}. A notable exception is Ref.~\cite{1367-2630-14-7-073050}, where bound states in interacting quantum walks are studied.

In the present paper we study a one-dimensional massive Fermionic cellular automaton with a four-Fermion on-site interaction. The main result consists in the complete analytical solution in the two-particle sector. The linear part of the evolution corresponds to a one-dimensional Dirac walk \cite{bisio2013dirac}, with an  interaction having the most general on-site, number-preserving form. The same kind of interaction characterizes the most studied integrable quantum systems \cite{lieb1968absence,PhysRevD.11.2088,Korepin1979,essler2005one} such as Hubbard's \cite{Hubbard238} and Thirring's \cite{THIRRING195891} models. For this reason we call the present model Thirring quantum cellular automaton.

Despite the similarities, the present cellular automaton differs from the above models mainly in the discreteness of time evolution. This feature produces non-trivial differences in the dynamical solutions of the model, in particular a wider spectrum of scattering states, and the existence of bound states for every value of the total momentum. As a consequence of the departure of the present discrete-time evolution from the usual Hamiltonian paradigm, we are not allowed to borrow the common Bethe ansatz technique straightforwardly.

We start defining a quantum walk for interacting particles on the
lattice $\mathbb{Z}$, assuming the particle statistics to be
Fermionic. First we introduce the walk $W$ for a free two-component
Fermionic field $\psi$ defined at any lattice point $x\in\mathbb{Z}$
and at any discrete time $t\in\mathbb{Z}$
\begin{equation}\nonumber
\begin{aligned}
&\psi(x,t+1)=W\psi(x,t),
\qquad \psi(x,t)=\begin{pmatrix}
      \psi_\uparrow(x,t)\\
      \psi_\downarrow (x,t)\end{pmatrix}
\\
&W=
    \begin{pmatrix}
      \nu T_x^\dag&-i\mu\\
      -i\mu & \nu T_x\end{pmatrix},\quad \nu,\mu >0,\: \nu^2+\mu^2=1,
\end{aligned}
\end{equation}
where $T_x$ is the translation operator $T_x \phi(x) = \phi(x+1)$ and
$\psi_\uparrow$ and $\psi_\downarrow$ denote the two components of the field.  In the
one-particle sector the above walk is a unitary operator $W$ over the
Hilbert space $\Hilb=\mathbb{C}^2 \otimes \ell^2(\mathbb{Z})$ for
which we will use the factorized orthonormal basis
$\ket{a}\ket{x}$, with $a\in\{\uparrow,\downarrow\}$.

Notice that the walk evolution is local, with the field at time $t$
and at site $x$ depending only on the field at sites $x\pm 1$
at time $t-1$ (first-neighbouring scheme). Moreover, since $W$
commutes with translations along the lattice, the walk can be
diagonalized in the momentum space. In the Fourier representation the
operator $W$ is expressed in terms of the momentum $p \in (-\pi,\pi]$
($ \ket{p}:= (2\pi)^{-1/2}\sum_x e^{-ipx} \ket{x}$) as follows
\begin{equation}\label{eq:free-qw} \begin{aligned}
    & W= \int \!\! \dd p \, {W}(p) \otimes \ketbra{p}{p}, \quad W(p)=
    \begin{pmatrix}
  \nu e^{ip} &-i\mu\\
  -i\mu & \nu e^{-ip}
\end{pmatrix}\\
 &   {W}(p) \bvec{v}^{s}_p = e^{- i  s\omega(p)} \bvec{v}^{s}_p,
\quad \bvec{v}^{s}_p :=
 \frac{1}{|N_s|} \begin{pmatrix}
  -i\mu\\
    g_s(p)
\end{pmatrix},\\
& \omega(p) := \mathrm{Arccos}(\nu \cos{p}),\qquad s\in\{+,-\},
\end{aligned}
\end{equation}
with $g_s(p)=-i (s \sin \omega(p) + \nu \sin p)$, 
$|N_s|^2 = \mu^2 + |g_s|^2$.

The function $\omega(p)$ is the walk
dispersion relation. In Ref.~\cite{Bisio2015244} it is shown that for small $p$ this
discrete dynamics recovers that of a free Dirac field of mass $\mu$. 

A $N$-particle walk can then be described taking
$\Hilb_N= \Hilb^{\otimes N}$ as the Hilbert space of the system and
$W_N:=W^{\otimes N}$ as the operator providing the evolution. Within
this scenario we introduce a coupling between particles defining the
dynamical time step via a unitary operator of the form
$W_NV_{int}$. Here we consider the following interacting term
\begin{align}\nonumber
&V_{int}:=V(\chi)= e^{i\chi n_\uparrow(x)n_\downarrow(x)},
\end{align}
where $n _a (x)=\psi^\dag_a(x) \psi_a(x)$ is the number operator at
site $x$ and with internal state $a\in\{\uparrow,\downarrow\}$. This
corresponds to an on-site coupling, namely the action of $V(\chi)$ is
non trivial if and only if two Fermions lie at the
same site of the lattice. Moreover, since $V(\chi)$ commutes with the
total number operator
$n=\sum_x(\psi_\uparrow^\dag(x) \psi_\uparrow
(x)+{\psi_\downarrow}^\dag (x) \psi_\downarrow(x))$, the dynamics
preserves the number of particles. This is the most general possible coupling with the above properties for the case of a Fermionic walk \cite{PhysRevB.44.12413}. In this paper we restrict to the
$N=2$ case and therefore we focus on the walk
\begin{align}\label{eq:automaton}
&U_2:=  W_2V_2(\chi),\qquad  V_2(\chi) := e^{i\chi\delta_{y,0}(1-\delta_{\alpha_1,\alpha_2})},
\end{align}
which is written in the center of mass basis $\ket{a_1,a_2}\ket{y}\ket{w}$ for
the two-particles Hilbert space
$\Hilb_2=\mathbb{C}^4\otimes\ell^2(\mathbb{Z})$, with
$a_1,a_2\in\{\uparrow,\downarrow\}$, $y= x_1-x_2$ and $w = x_1+x_2$
the relative and the center of mass coordinate respectively. Denoting
by $k= \tfrac{1}{2}(p_1-p_2)$ the (half) relative momentum and by
$p = \tfrac{1}{2}(p_1+p_2)$ the (half) total momentum, the free walk
$W_2$ in the momentum representation is written as follows
\begin{align}\label{eq:free-qw-2p}
    &{W}_2 = \int \!\! \dd k \dd p \,  {W}_2(p,k) \otimes \ketbra{k}{k} \otimes \ketbra{p}{p},\\
&W_2(p,k) \bvec{v}^{sr}_{p,k},=
e^{-i\omega_{sr}(p,k)}\bvec{v}^{sr}_{p,k},\quad \bvec{v}^{sr}_{p,k}:=
  \bvec{v}^{s}_{p+k} \otimes \bvec{v}^{r}_{p-k},
  \nonumber\\
&\omega_{sr}(p,k):=s\omega(p+k)+r\omega(p-k),\quad s,r\in\{+,-\},
\nonumber
\end{align}
where the eigenvectors of ${W}_2(p,k) := W(p+k) \otimes W(p-k)$ are
easily computed as the tensor product of the single-particle ones in
Eq.~\eqref{eq:free-qw}.

Since the interacting dynamics $U_2$ commutes with translation $T_w$ in the center of
mass coordinate $w$, it is convenient to write the walk in the
hybrid basis $\ket{a_1,a_2}\ket{y} \ket{p}$, in the following block-diagonal form
\begin{equation}
\begin{aligned}
  & U_2 = \int \dd p \, U_2(\chi,p) \otimes \ketbra{p}{p}, \quad U_2(\chi,p):= W_2(p)  \tilde{V}_2(\chi), \nonumber\\
  &W_2(p):=\mu\nu
 \begin{pmatrix}
    \frac{\nu}{\mu} e^{i2p}   & -i e^{ip} T_y&
 -i  e^{ip}  T_y^\dag  & -\frac{\mu}{\nu} \\
-i e^{ip} T_y & \frac{\nu}{\mu} T_y^2 &
 -\frac{\mu}{\nu} & -i e^{-ip} T_y \\
-i e^{ip} T_y^\dag  & -\frac{\mu}{\nu} &
\frac{\nu}{\mu} {T_y^\dag}^2 & -i e^{-ip}  T_y^\dag \\
    -\frac{\mu}{\nu}  & -i e^{-ip} T_y &
-i  e^{-ip}T_y^\dag & \frac{\nu}{\mu} e^{-i2p}
  \end{pmatrix}\!, \nonumber\\
&\tilde{V}_2(\chi) :=
  \begin{pmatrix}
   I &0&0&0 \\
  0 &e^{i\chi \delta_{y,0}}I&0&0 \\
  0  &0&e^{i\chi \delta_{y,0}}I&0 \\
  0  &0&0& I
  \end{pmatrix},
\end{aligned}
\end{equation}
with $T_y$ the translation in the relative coordinate $y$.

The first step is to solve the linear
difference equation
\begin{align}\label{eq:eigenvaluemothereq}
 U_2(\chi,p) \bvec{f}_{p,\omega,\chi}   = e^{i \omega}
  \bvec{f}_{p,\omega,\chi},   \\
   \bvec{f}_{p,\omega,\chi} : \mathbb Z \to \mathbb{C}^4, \quad  \omega
  \in \mathbb{C},\nonumber
\end{align}
for any possible value of $ \chi $ and $ p $.  Among all the possible
solutions of Equation~\eqref{eq:eigenvaluemothereq} we will then
choose those ones which are eigenvectors (or generalized eigenvectors)
of $ U_2(\chi,p)$ considered as an operator on the Hilbert space
$\mathbb{C}^4\otimes \ell^2(\mathbb Z)$.  Since the interacting
particles are Fermions, we are only interested in the solutions that
are antisymmetric under the exchange of the two particles, i.e.
\begin{align}\nonumber
  \bvec{f}_{p,\omega,\chi}  (y)  = -E \bvec{f}_{p,\omega,\chi}  (- y),
\end{align}
where $E$ is represented as
$E=\tfrac{1}{2}\sum_{i=0}^{3}\sigma_i\otimes\sigma_i$ (with
$\sigma_0=I$, and $\sigma_{i}$, $i=1,2,3$, the Pauli matrices).

In the following, in order to lighten the notation, we will omit
the explicit dependence of the solutions from
$p,\omega,\chi$ and we will write
$\bvec{f}  (y)  $ for
$\bvec{f}_{p,\omega,\chi}  (y)  $.

\begin{figure}[tb]
  \begin{center}
    \includegraphics[width=.8\columnwidth]{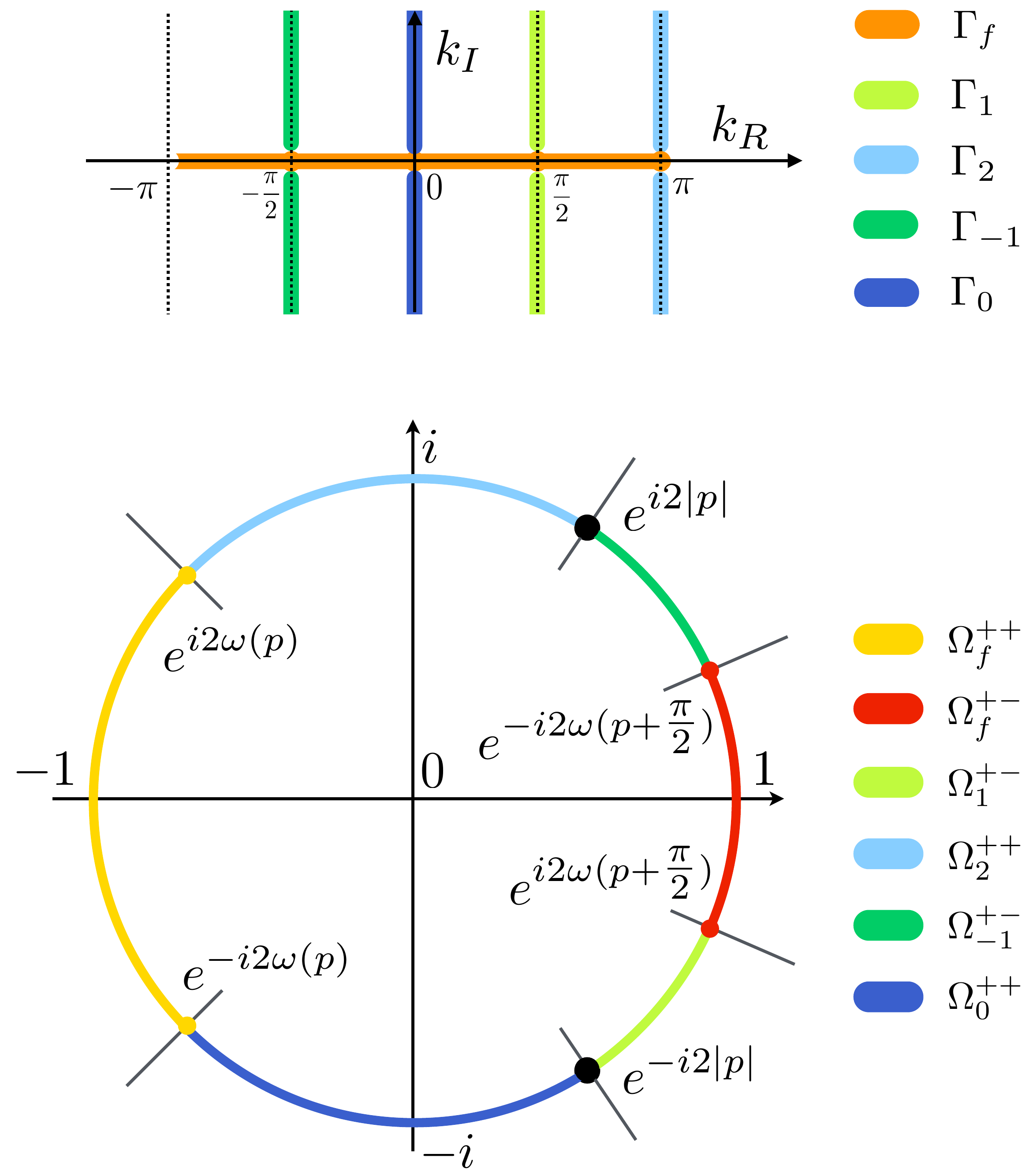}
  \end{center}
  \caption{(Colors online) Top: $k_R$ and $k_I$ represent the
    real and the imaginary part of the relative momentum $k$ in the two-fermion state. 
    The highlighted regions collect the values of $k\in\mathbb C$ providing 
    a real value of the quasi-energy $\omega$.  Bottom: the disjoint subregions 
    of the unit circle are the images under 
    $k\mapsto e^{i\omega_{rs}(p,k)}$ of the 
    disjoint regions in the top figure, 
    for fixed values of the total momentum $p=0.55$ and mass
    $\mu=0.8$.  $\Omega_f$ coincides with the 
    continuous spectrum of $U_2(\chi,p)$ (see
    Eq.~\eqref{eq:automaton}).
     The discrete spectrum lies in the
    other regions, and 
    for a fixed value of the coupling constant $\chi$ it consists of a single point.
    Varying the value of $\chi$ the unit circle is covered, 
    and the boundary points of the arcs depend on $p$.
    \label{f:setsfig}}
\end{figure}

Since the interacting term acts only at the origin, for $y > 0 $
Eq.~\eqref{eq:eigenvaluemothereq} becomes a linear recurrence
relation with constant coefficient whose most general solutions
\cite{cull2005difference} are of two forms: $\bvec f_\infty(y)$ or $\bvec f(y)$, given by
\begin{align}
\label{eq:generalsolution}
  &{\bvec{f}_{\infty}} (y)  =
    ( {\zeta}_\infty , 0 , 0,  {\zeta'}_\infty   ) ^T 
    \delta_{y,1} \quad  {\zeta}_\infty,  {\zeta'}_\infty \in
    \mathbb{C}  \quad y>0, \\
\label{eq:generalsolutiondegenerate}
  &\bvec{f}  (y)  =
    \sum_{s,r=\pm} \int_{\mathsf{S}}  \!  \dd {k}
    \,    e^{- i k y}  g_\omega(s,r,k) \bvec{v}_k^{sr}    \qquad y>0 \\
  & k=k_R+ik_I, \:\: \mathsf{S} := \{k\in\mathbb C\mid k_R \in
    (-\pi,\pi]  \}, \nonumber \\
  & g_\omega(s,r,k) \in
    \mathbb{C} \mbox{ s.t. }   e^{-i \omega} \neq e^{-i \omega_{sr}(p,k)} \Rightarrow
    g_\omega (s,r,k)=0, \nonumber
\end{align}
where the function $\omega_{rs}(p,k)$ and the vectors
$\bvec{v}_k^{sr}:=\bvec{v}_{p,k}^{sr}$ have been defined in
Eq.~\eqref{eq:free-qw-2p}, and for complex argument $\mathrm{Arccos}$ is the principal value of the arccosine function.

Let us first consider the functions given by
Eq.~\eqref{eq:generalsolution}.
A necessary condition for a function obeying
Eq.~\eqref{eq:generalsolution} to be a (proper or improper)
eigenvector of $U_2(\chi,p)$ is that
$\omega_{sr}(p,k)\in\mathbb{R}$. In order to analyse this condition, it is useful to introduce the following sets (see
Fig.~\ref{f:setsfig}):
\begin{align}
  &\begin{aligned}
 &   \Gamma_f := \{ k \in \mathsf{S} | k_I =0 \},
    \\
&  \Gamma_{z}:=
    \{ k \in \mathsf{S} | k_R = z \tfrac{\pi}{2} \}, \quad z = 0,\pm 1
    ,2,
  \end{aligned} \label{eq:Gammasets2}
 \\
  &
    \begin{aligned}
  &\mathrm{\Omega}^{sr}_{f} :=
  \{   \exp( - i \omega_{sr}(p,k)) \,|\, k \in \Gamma_{f}\},
  \\
    &\mathrm{\Omega}^{sr}_{z} :=
\{   \exp( - i \omega_{sr}(p,k))  \,|\, k \in \Gamma_z,\ (-1)^z=sr \}.
    \end{aligned} \label{eq:Omegasets2}
\end{align}
Reminding that $\omega(x + \pi ) = \pi - \omega(x) $,
one can verify that
\begin{align}
  \label{eq:omegaidentities}
  \begin{aligned}
    \mathrm{\Omega}^{++}_{f} = \mathrm{\Omega}^{--}_{f},\quad
    \mathrm{\Omega}^{+-}_{f} = \mathrm{\Omega}^{-+}_{f},\quad
    \mathrm{\Omega}^{++}_{0} = \mathrm{\Omega}^{--}_{2},\\
    \mathrm{\Omega}^{++}_{\pi} = \mathrm{\Omega}^{--}_{0},\quad
    \mathrm{\Omega}^{+-}_{-1} =
    \mathrm{\Omega}^{-+}_{1},\quad
    \mathrm{\Omega}^{+-}_{1} =
    \mathrm{\Omega}^{-+}_{-1}.
  \end{aligned}
\end{align}
The following technical result, proved in Appendix~\ref{sec:proof-lemma-refl}, marks the first important difference from the Hamiltonian integrable models, relying in the degeneracy of two-particle levels. The degeneracy in the Hamiltonian case is two, corresponding to the intuitive one-dimensional picture where either a classical elastic bounce or a quantum tunnelling where the particles cross each other occur. On the other hand, in the discrete case the degeneration is four, allowing also for scattering events where hopping to a distant region in the Brillouin zone can occur. This phenomenon is due to periodicity of the quasi-energy spectrum, which is where the failure of the Bethe ansatz lurks.

\begin{lemma} \label{lem:setsandsol}
  Let $\omega_{sr}(p,k)$ be defined as in Eq.~\eqref{eq:free-qw-2p} and let us assume
   $p \neq z\frac{\pi}{2}$ ($z \in
   \mathbb{Z}$).
   Then we have:
   
   \begin{enumerate}

   \Item\label{i:1}
     \begin{align}
     \label{eq:zones}
     \begin{aligned}
       \omega_{\pm\pm}(p,k) \in \mathbb{R} &\implies k \in \Gamma_{\mathrm{f}} \cup
     \Gamma_{0} \cup
     \Gamma_{2} \\
         \omega_{\pm\mp}(p,k) \in \mathbb{R} &\implies k \in
         \Gamma_{\mathrm{f}} \cup
          \Gamma_{-1} \cup
     \Gamma_{1}
   \end{aligned}
            \end{align}
            
          \item \label{i:2}The six sets
            $\mathrm{\Omega}^{++}_{f} ,
    \mathrm{\Omega}^{+-}_{f} ,
    \mathrm{\Omega}^{++}_{0} ,
    \mathrm{\Omega}^{++}_{2} ,
    \mathrm{\Omega}^{+-}_{-1} ,
    \mathrm{\Omega}^{+-}_{1} $
    are disjoint and their union is the whole unit circle except for the
    points $e^{ \pm i 2p }$.
    \item\label{i:3} For any $\omega \in \mathbb{R}$ such that $e^{-i \omega} \neq e^{\pm
        i 2p}$ the equation
            $ e^{-i \omega} = e^{-i \omega_{sr}(p,k)} $ has four
            distinct solutions. If the triple $(+,+,k)$ is a solution
            then also
            $(+,+,-k)$, $(-,-,\pi-k)$ and $(-,-,k-\pi)$ are solutions.
 If      the triple   $(+,-,k)$ is a solution
          then
            $(-,+,-k)$, $(+,-,\pi-k)$ and $(-,+,k-\pi)$ are solutions.
   \end{enumerate}
\end{lemma}
By Lemma~\ref{lem:setsandsol},
Eq.~\eqref{eq:generalsolution} yields two classes of
solutions:
\begin{align}
 & \begin{aligned}
  &\bvec{f}_k^{+}(y) \mbox{ with } k \in  \Gamma_f  \cup \Gamma_0
  \cup \Gamma_2,\\
   &\bvec{f}_k^{-}(y) \mbox{ with } k \in  \Gamma_f  \cup \Gamma_{-1} \cup \Gamma_1  ,
 \end{aligned}
\label{eq:solclass}
  \\
& \begin{aligned}
   \bvec{f}_k^{\pm}(y) =
     \begin{cases}
     \begin{array}{l}
       [\alpha_{\pm} \bvec{v}^{+ \pm}_k -
       (-1)^y \delta_{\pm} \bvec{{v}}^{- \mp}_{k-\pi}]
       e^{-iyk} +\\
\; \;-[\beta_{\pm} \bvec{v}^{\pm +}_{-k} -(-1)^y \gamma_{\pm}
       \bvec{{v}}^{\mp -}_{\pi-k}   ]e^{iyk}
 \end{array}  &  y>0
     \\
 (0,\eta_{\pm},-\eta_{\pm},0)^T & y = 0
     \\
   \mbox{ antisymmetrized } & y<0,
  \end{cases}
 \end{aligned}
\nonumber
\end{align}
where $\alpha_{\pm}, \beta_{\pm}, \dots $, are complex coefficients
which depend on $p,k,m,\chi$.  We now determine these
coefficients by requiring that Eq.~\eqref{eq:eigenvaluemothereq} is
satisfied. Because of the locality of the evolution, this constraint
needs to be verified only for $y=0,1,2$.  A tedious
albeit straightforward calculation allows one to bring Eq.~\eqref{eq:solclass}
into the following form after suitable reparametrization
\begin{align}
  &\bvec{f}_k^{\pm}(y)= c_1 \bvec{f}_k^{\pm , f}(y) + c_2
  \bvec{f}_k^{\pm, i}(y),
\label{eq:solclass2}
  \\
&\begin{aligned}
\bvec{f}_k^{\pm , f }(y) =&
       [\bvec{v}^{+ \pm}_k +
       (-1)^y \bvec{{v}}^{- \mp}_{k-\pi}]
       e^{-iyk} +\nonumber\\
&-[\bvec{v}^{\pm +}_{-k} +(-1)^y
       \bvec{{v}}^{\mp -}_{\pi-k}   ]e^{iyk},
\end{aligned}
  \nonumber\\
  &
    \bvec{f}_k^{\pm , i}(y) =
    \begin{cases}
      \begin{array}{l}
      e^{-i \delta_{y,0} \chi} \{ [\bvec{v}^{+ \pm}_k -
       (-1)^y \bvec{{v}}^{- \mp}_{k-\pi}]
       e^{-iyk} + \\
 \, -T_{\pm} [\bvec{v}^{\pm +}_{-k} -(-1)^y
       \bvec{{v}}^{\mp -}_{\pi-k}   ]e^{iyk} \}
      \end{array}
      & y \geq 0\\
% (0, ????,  ????? ,0 )^T
%             & y = 0\\
      \mbox{antisymmetrized}
            & y < 0,
    \end{cases}
              \nonumber\\
  &T_{\pm}:=
  \frac{
  g_{+} (p+k) + e^{-i \chi} g_{\pm}(p-k)
  }
  {
    g_{\pm} (p-k) + e^{-i \chi} g_{+}(p+k)
    },
    \qquad
    c_1,c_2 \in \mathbb{C}.
    \nonumber
\end{align}
The next step of the analysis is to identify, among the set of
functions of Eq.~\eqref{eq:solclass2},
those ones which correspond to eigenvectors or generalized eigenvectors
of $U_2(\chi,p)$.

For $k \in \Gamma_f $, Eq.~\eqref{eq:solclass2} gives the generalized
eigenvector of $U_2(\chi,p)$ corresponding to the continuous spectrum
$\sigma_c = \Omega^{++}_f \cup \Omega^{+-}_f $. Since
$U_2-W_2$ is a finite rank operator, the continuous
spectrum of the two-particle interacting case is the same as that of the free
walk (see Theorem IV 5.35 of Ref.\cite{kato2013perturbation}).  From
Eq.~\eqref{eq:solclass2} we have that the solutions of the kind
$\bvec{f}_k^{\pm , f}$ are generalized eigenvectors of the free theory
which are also generalized eigenvectors of the interacting
theory. This is easily understood since
$\bvec{f}_k^{\pm , f}(0) = \bvec{0}$, and therefore those
eigensolutions do not feel the presence of the interaction---which is localized at $y=0$.  On the other hand, we can
interpret the solution of the kind $\bvec{f}_k^{\pm , i}$ as a
scattering of plane waves with the $T_{\pm}$ playing the role of
transmission coefficients.

For $ k \not \in \Gamma_f $, necessary conditions for
$\bvec{f}_k^{\pm}$ to be a (proper or generalized) eigenvector of
$U_2(\chi,p)$ are that $ k_I = \Im(k)<0 $, $c_1=0$, $c_2 \neq 0$ and
$T_\pm=0$ (otherwise $\bvec{f}_k^{\pm}$ is exponentially divergent).
In appendix \ref{a:lemma-boundstates} we prove the following result:

\begin{lemma}\label{lem:boundstates}
  Let $T_{\pm}$ defined as in Eq.~\eqref{eq:solclass2}
  and let us assume $p \neq z \tfrac{\pi}{2}$.
  If $e^{i\chi} \not \in \{ e^{\pm i 2 p},1,-1 \}$,
  then there exists a unique $k \in \Gamma_0\cup\Gamma_{-1}\cup
  \Gamma_1 \cup \Gamma_2 $ with $k_I<0$
  such that  either $T_{+} =0$ or $T_{-} =0$.
  On the other hand, if
  $e^{i\chi} \in  \{ e^{\pm i 2 p},1,-1 \}$ then
  $T_{+} \neq 0$ and $T_{-} \neq 0$ for all $k \in \Gamma_0\cup\Gamma_{-1}\cup
  \Gamma_1 \cup \Gamma_2 $ with $k_I<0$.
\end{lemma}
The above result tells us that, for $e^{i\chi} \not\in\{ e^{\pm i 2 p},1,-1\}$, 
the two-particles interacting evolution $U_2(\chi,p)$ has one proper eigenvector whose corresponding
eigenvalue constitutes the discrete spectrum of $U_2(\chi,p)$.
This eigenstate is easily interpreted as a bound state of two
particles.

We now consider the 
functions given by Eq.~\eqref{eq:generalsolutiondegenerate}
which lead to the  antisymmetric functions
\begin{align}
  \label{eq:degeneratesolution}
 &{\bvec{f}_\infty}  (y)  =
    \left \{
    \begin{aligned}
   & ( \zeta_\infty , 0 , 0 , \zeta'_\infty   )^T  \delta_{y,1}& y>0,\\
    & (0, {\eta}_\infty, - {\eta}_\infty ,0)^T   & y=0, \\
    & ( -\zeta_\infty , 0 , 0 , -\zeta'_\infty   )^T \delta_{y,-1}& y<0. 
  \end{aligned}
\right. 
\end{align}
Imposing condition~\eqref{eq:eigenvaluemothereq}, we obtain the following solutions
\begin{align}
  \label{eq:degeneratesolution2}
 &{\bvec{f}_{\pm \infty}}  (y)  =
    \left \{
    \begin{aligned}
   & i  e^{\pm i p}( -\tfrac{1\pm 1}{2}, 0 , 0 , \tfrac{-1\pm 1}{2} )^T  \delta_{y,1}& y>0,\\
    & (0,   \tfrac{\mu}{\nu}, - \tfrac{\mu}{\nu} ,0)^T   & y=0, \\
    &i  e^{\pm i p}( \tfrac{1\pm 1}{2}, 0 , 0 ,- \tfrac{-1\pm 1}{2} )^T  \delta_{y,-1}& y<0, 
  \end{aligned}
\right. \nonumber\\
& U_2(\chi,p) \bvec{f}_{ \pm \infty}  = e^{ \pm i 2p} \bvec{f}_{\pm
  \infty} \mbox{ for } e^{i\chi} =  e^{\pm i2p}.
\end{align}
Eq.~\eqref{eq:degeneratesolution2} provides the proper
eigenstate of $ U_2(\chi,p)$ for the cases
$ e^{i\chi} = e^{\pm i2p}$ which were missing in
Lemma~\ref{lem:boundstates}.

\begin{figure}[th]
\includegraphics[width=0.8\columnwidth]{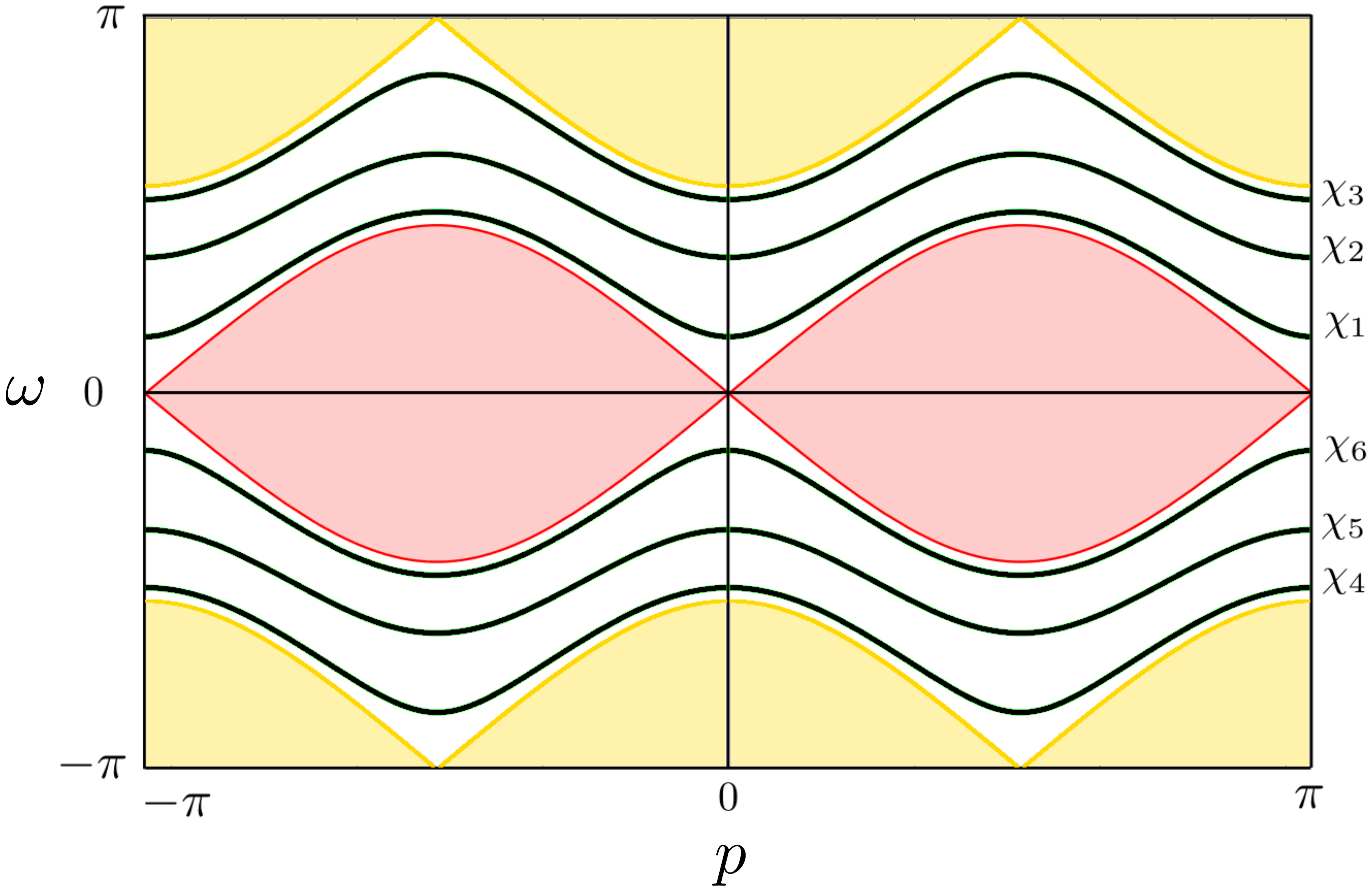}
\caption{Spectrum of the 2 particle automaton of
  Eq.~\eqref{eq:automaton}: In red and yellow are depicted the
  continuous spectrum bands; in black the discrete band for different
  values of the coupling: $\chi_1 = -\frac{\pi}{5}$,
  $\chi_2 = -\frac{\pi}{2}$, $\chi_3 = -\frac{4\pi}{5}$,
  $\chi_4 = \frac{4\pi}{5}$, $\chi_5 = \frac{\pi}{2}$,
  $\chi_6 = \frac{\pi}{5}$.  }\label{f:fullspectrum}
\end{figure}

In Fig.~\ref{f:fullspectrum} we plot the spectrum of
$U_2(\chi,p)$ as a function of $p$ for different values of the $\chi$.
We can then write,  for
$ p \neq z \tfrac{\pi}{2} $, the spectral
resolution of  $U_2(\chi,p)$, i.e.
\begin{align*}
 U_2(\chi,p) =      &\sum_{\substack{s=\pm,\\ j =f,i}} \! \int_{-\pi}^{\pi}
\!\!\! \!\! \!
\dd k  \,
e^{-i\omega_{+s}(p,k)}
 \ketbra{\phi_{p,\chi}^{s,j}(k)}{\phi_{p,\chi}^{s,j}(k)}
   + \nonumber \\
&+ e^{-i
  \tilde{\omega}}\ketbra{\varphi_{p,\chi}}{\varphi_{p,\chi}},
\end{align*}
where we defined
  \begin{align*}
 &   \begin{aligned}
 & \braket{y}{\phi_{p,\chi}^{s,j}(k)} := N_{p,\chi,s,j,k}  \, \mathbf{f}_k^{s,j}(y),\\
&
  \braket{y}{\varphi_{p,\chi}} := 
                  \begin{cases}
                            M_{p,\chi,\tilde{k}}      \,   \mathbf{f}_{\tilde{k}}^{+,i}(y), &
e^{i\chi} \neq e^{\pm i 2 p},\      T_+(\tilde{k}) =0, \\
M_{p,\chi,\tilde{k}} \, \mathbf{f}_{\tilde{k}}^{-,i}(y),&
e^{i\chi} \neq e^{\pm i 2 p},\   T_-(\tilde{k}) =0, \\
M^{\pm }_{p}      \,   \mathbf{f}_{\pm \infty}(y), &
e^{i\chi} = e^{\pm i 2 p},
\end{cases}\\
& \tilde{\omega}:=
\begin{cases}
                            \omega_{++}(p,\tilde{k}),&
e^{i\chi} \neq e^{\pm i 2 p}, \   T_+(\tilde{k}) =0, \\
 \omega_{+-}(p,\tilde{k}),&
e^{i\chi} \neq e^{\pm i 2 p},\  T_-(\tilde{k}) =0, \\
\pm 2p, &
e^{i\chi} = e^{\pm i 2 p},
\end{cases}
                \end{aligned}
  \end{align*}
and $N, M$ are normalization factors
such that
  \begin{align*}
    &\braket{\phi_{p,\chi}^{s,j}(k)}{\phi_{p,\chi}^{s',j'}(k')}=
    \delta_{s,s'} \delta_{j,j'} \delta(k-k') \\
&    \braket{\varphi_{p,\chi}}{\varphi_{p,\chi}} =  1 
 \end{align*}

We conclude our analysis with the discussion of the cases
$p = z \tfrac{\pi}{2}$ starting from $p=0$.  We have
$\omega_{\pm \pm}(0, k) = \pm2\omega(k) $, with $\omega(k) \in (-\pi,\pi]$ and $\omega(k)\neq0$,
iff $k\in \Gamma_f \cup\Gamma_0\cup \Gamma_2$. On the other hand
$\omega_{\pm \mp}(0,k) = 0 $ for all $k\in \mathbb{C}$, and thus $\omega_{\pm \mp}(0,k)\neq \omega_{\pm \pm}(0,k')$ for all values of $k,k'$. Therefore the
previous analysis still holds for $e^{- i \omega} \neq 1 $ and, by
setting $p=0$, the solutions ${\bf{f}} _k^{+}$ of
Eq.~\eqref{eq:solclass2} are (proper and improper) eigenvectors of
$U_2(\chi,0)$.  Thus, the spectrum of $U_2(\chi,0)$ decomposes into a continuous
spectrum, which is the arc of the unit circle wich contains $-1$ and
has $ e^{\pm 2i \omega(0)}$ as extremes, and a point spectrum made of
two distinct points: $e^{-i 2\omega(\tilde{k})}$ (where $\tilde{k}$ is the solution
of $T_+=0$ when $p=0$) and $1$.  Since $U_2(\chi,0)$ is unitary, if $e^{- i \omega}$ 
belongs to the point spectrum then it is a proper eigenvalue of $U_2(\chi,0)$.  Let us denote
with $P_0^{-}$ the projection on the eigenspace of the eigenvalue $1$, and by $P^-_p$ the following projection
\begin{align*}
P^{s}_p:=\sum_{ j =f,i}
\int_{-\pi}^{\pi}
\!\! \!\!\!
\dd k  \,
\ketbra{\phi_{p,\chi}^{s,j}(k)}{\phi_{p,\chi}^{s,j}(k)}.
\end{align*}
%  Then there exists a (possibly
% $\mathcal{H}_{0}^{(-)}$ 
%infinite) orthonormal set of proper eigenvectors
% $\ket{\psi_n} \in \mathbb{C} \otimes \ell^2(\mathbb{Z}) $ which is a
% basis for $\mathcal{H}_{0}^{-}$ and such that
% $P_0^{-}=\sum_n\ketbra{\psi_n}{\psi_n}$.
%which can be characterised as the limit of $P^-_p$ for $p\to0$ in the 
%operator strong norm $ \|\cdot \| $, where
%\begin{align*}
%P^{s}_p:=\sum_{ j =f,i}
%\int_{-\pi}^{\pi}
%\!\! \!\!\!
%\dd k  \,
%\ketbra{\phi_{p,\chi}^{s,j}(k)}{\phi_{p,\chi}^{s,j}(k)}.
%\end{align*}
%Indeed,
Now, since $\lim_{p \to 0} \| U_2(\chi,p) - U_2(\chi,0) \| = 0 $, and $1$ is a separated part of the spectrum of $U_2(\chi,0)$,
then $\lim_{p \to 0} \| P^{-}_p - P_0^{-} \| = 0$ 
 (see Theorem IV 3.16 of Ref.\cite{kato2013perturbation}).
 We have then that
 \begin{align*}
   \begin{aligned}
 &  P^-_0= \sum_{\substack{n\in\mathbb{Z}\\ j = f,i}}\ket{\psi_0^{-}(n,j)}\bra{\psi_0^{-}(n,j)} ,\\
&\ket{   \psi_0^{-}(n,j)} := \int_{-\pi}^\pi  g_{n}(k)\ket{\phi_{0,\chi}^{-,j}(k)}
   \end{aligned}
    \end{align*}
    where $g_{n}(k)$ is an orthonormal basis for
    $L^2(-\pi, \pi] $.
    The cases $p= \pi, \pm \tfrac{\pi}{2} $
    can be analysed in the same way. The eigenspace corresponding to the eigenvalue $1$ is thus a separable Hilbert space of stationary bound states. This result marks an important departure from the behaviour of analogous Hamiltonian models. Remarkably, it occurs even in the non-interacting case $\chi=0$.

The diagonalisation of $U_2(\chi,p) $
%is then complete for any value of $p$ and it is 
summarized by the
following proposition
\begin{proposition}
Let $U_2(\chi,p)$
be defined as in Equation~\eqref{eq:free-qw-2p}.
Then its spectral
resolution is
\begin{align*}
  & U_2(\chi,p) =\\
&= \left \{
  \begin{aligned}
    &   \sum_{\substack{s=\pm,\\ j =f,i}}
U^{s,j}_{p,\chi} + e^{-i  \tilde{\omega}} P_{p,\chi} &\; \;&  p \neq  z\tfrac{\pi}{2}, \\
    &   \sum_{ j =f,i}
    U^{+,j}_{z \pi,\chi} + P_{z \pi}^{-}+
    e^{-i  \tilde{\omega}} P_{z \pi,\chi} &\; \;& p =  0, \pi, \\
    % U_2^{s,j}(p,\chi) + e^{-i  \tilde{\omega}} P_{p,\chi}(\tilde{k}) \qquad  p \neq  z\tfrac{\pi}{2} \\
    &   \sum_{ j =f,i}
    U^{-,j}_{\pm \tfrac{\pi}{2},\chi} - P_{\pm \tfrac{\pi}{2}}^{+}+
    e^{-i  \tilde{\omega}} P_{\pm \tfrac{\pi}{2},\chi}  &\; \;& p =  \pm \tfrac{\pi}{2},
\end{aligned}
\right . \nonumber
    \\
  & U^{s,j}_{p,\chi} := \int_{-\pi}^{\pi}
\!\!\! \!\! \dd
  k \,
    e^{-i\omega_{+s}(p,k)}
    \ketbra{\phi_{p,\chi}^{s,j}(k)}{\phi_{p,\chi}^{s,j}(k)},
  \nonumber\\
    & P_{p,\chi} =
      \ketbra{\varphi_{p,\chi}}{\varphi_{p,\chi}}.
      \nonumber
%   \dd P_{p,\chi}^{(s,j)}(k)    \\
%   &+ e^{-i
%   \tilde{\omega}} P_{p,\chi}(\tilde{k}),  \qquad  p \neq  z\tfrac{\pi}{2} \\
% & \sum_{\substack{s=\pm,\\ j =f,i}}\int_{-\pi}^{\pi}
% \!\!\! \!\!
%   e^{-i\omega_{+s}(p,k)}
%   \dd P_{p,\chi}^{(s,j)}(k)    + \\
%  & e^{-i \tilde{\omega}} P_{p,\chi}(\tilde{k})  \quad p \neq  z\tfrac{\pi}{2}
% \end{aligned}
% \right .         \\
%     \left \{
%   \begin{aligned}
%   &   \sum_{\substack{s=\pm,\\ j =f,i}}\int_{-\pi}^{\pi}
% \!\!\! \!\!
%   e^{-i\omega_{+s}(p,k)}
%   \dd P_{p,\chi}^{(s,j)}(k)    \\
%   &+ e^{-i
%   \tilde{\omega}} P_{p,\chi}(\tilde{k}),  \qquad  p \neq  z\tfrac{\pi}{2} \\
% & \sum_{\substack{s=\pm,\\ j =f,i}}\int_{-\pi}^{\pi}
% \!\!\! \!\!
%   e^{-i\omega_{+s}(p,k)}
%   \dd P_{p,\chi}^{(s,j)}(k)    + \\
%  & e^{-i \tilde{\omega}} P_{p,\chi}(\tilde{k})  \quad p \neq  z\tfrac{\pi}{2}
% \end{aligned}
% \right .         \\
%   & \sum_{\substack{s=\pm,\\ j =f,i}}\int_{-\pi}^{\pi}
% \!\!\! \!\!
% \dd k  \,
%   e^{-i\omega_{+s}(p,k)}
% \dd P_{p,\chi}^{(s,j)}(k)
%    + \nonumber\\
% & + e^{-i
%   \tilde{\omega}} P_{p,\chi}(\tilde{k})
%   \quad \mbox{for } p \neq  z\tfrac{\pi}{2} \\
%  & \dd P_{p,\chi}^{(s,j)}(k)   =
%   \ketbra{\phi_{p,\chi}^{(s,j)}(k)}{\phi_{p,\chi}^{(s,j)}(k)}  \dd k
%   \\
%                 &
% P_{p,\chi}(\tilde{k})  :=                 \ketbra{\varphi_{p,\chi}(\tilde{k})}{\varphi_{p,\chi}(\tilde{k})}
\end{align*}
\end{proposition}
We diagonalized an on-site interacting fermionic cellular automaton in the two-particle sector. Differently from analogous Hamiltonian models, (i) bound states exist for every value of the total momentum, (ii) there are four classes of scattering solutions instead of two, (iii) the bound states exist also in the free case. 
\bibliographystyle{apsrev4-1}
\bibliography{bibliography}

\newpage

\appendix

\section{Proof of Lemma~\ref{lem:setsandsol}}\label{sec:proof-lemma-refl}

\subsection{Proof of item~\ref{i:1}} \label{sec:proofrealomega}
Let us define
$ \hat{\omega}_{\pm}+i \tilde{\omega }_{\pm} := \omega(p\pm k)$.  Since
$\omega (z^*)=\omega^*(z)$, we have that both
$\hat{\omega}_\pm$ and $\tilde{\omega}_\pm$ are real.
Then we have
\begin{align}
&  \Im(\omega_{rs}(p,k))=0 \iff
 r \tilde{\omega }_{+} = -s \tilde{\omega}_-
  \implies \nonumber \\
&\cosh \tilde{\omega }_{+} = \cosh \tilde{\omega }_{-} =: \cosh\tilde{\omega}.
    \label{eq:realomega}
\end{align}
Reminding that $\cos\omega (p\pm k)=\nu\cos (p\pm k)$,
Eq.~\eqref{eq:realomega} implies that
\begin{align}
  \label{eq:1}
  \begin{aligned}
    \cos^2\hat{\omega}_{\pm} \cosh^2\tilde{\omega}&=
    \nu^2 \cos^2 (p\pm k_R) \cosh^2k_I\\
      \sin^2\hat{\omega}_{\pm} \sinh^2\tilde{\omega}&=
    \nu^2 \sin^2 (p\pm k_R) \sinh^2k_I.
  \end{aligned}
\end{align}
From the above relations we find that
\begin{align}\label{eq:cond33}
\cos^2(p \pm k_R)\frac{\cosh^2 k_I}{\cosh^2\tilde\omega}+\sin^2 (p \pm
  k)\frac{\sinh^2k_I}{\sinh^2\tilde\omega}=
\frac{1}{\nu^2}
\end{align}
which gives
\begin{align}\nonumber
  [\sin^2(p + k_R )-\sin^2( p - k_R )]\left(
  \frac{\sinh^2  k_I}{\sinh^2\tilde\omega}-\frac{\cosh^2  k_I}{\cosh^2\tilde\omega}
  \right)=0.
\end{align}
Now, since
$\frac{\sinh^2k_I}{\sinh^2\tilde\omega}- \frac{\cosh^2  k_I}{\cosh^2\tilde\omega}=0$
implies
$\frac{\sinh^2k_I}{\sinh^2\tilde\omega}= \frac{\cosh^2  k_I}{\cosh^2\tilde\omega}=1$,
which is not compatible with Eq.~\eqref{eq:cond33}, it must be
$\sin^2 (p + k_R)=\sin^2 (p-k_R)$, which gives
\begin{align*}
%\label{e:constraintappend}
  \begin{aligned}
      k_R = \frac{z}{2} \pi  \lor
p =  \frac{z}{2}\pi,
  \end{aligned}
\qquad z \in \mathbb{Z}.
\end{align*}
By explicit computation one obtains
\begin{align}
  \begin{aligned}
&\Im(\omega_{\pm\pm}(p,k))=0 \land k_I \neq 0
 \implies
    k_R = 0, \pi  \lor
p =  \pm \frac{\pi}{2},\\
&\Im(\omega_{\pm\mp}(p,k))=0 \land k_I \neq 0
 \implies
    k_R = \pm \frac{\pi}{2},  \lor
p = 0,\pi,
  \end{aligned}
\nonumber
\end{align}
which proves the first item of Lemma~\ref{lem:setsandsol}.

\subsection{Proof of item~\ref{i:2}}
Let us consider the case in which
$p \in (0,\tfrac{\pi}{2}) $.
The function
$k \mapsto \omega_{++}(p,k)$
is smooth and periodic with period $2\pi$ and
therefore it ranges beetween its maximum and minimal values.
The maximum and minimum values are found by setting
$\partial_k \omega_{++}(p,k) =0$.
By explicit computation one obtains
\begin{align*}
  \frac{\sin(p+k)}{\sqrt{1-\nu^2\cos^2(p+k)  }}
  =
  \frac{\sin(p-k)}{\sqrt{1-\nu^2\cos^2(p-k)  }},
\end{align*}
which implies, for $p \neq z \tfrac{\pi}{2}$, that
$k =0,\pi$.
we have than that
$\omega_{++}(p,k)$ ranges between
$2\omega(p)$ and $2\pi - 2\omega(p)$.
By noticing that $\omega_{++}(p,\tfrac{\pi}{2}) = \pi $
we have that $\Omega^{++}_f$ is the arc which connects
$e^{i 2 \omega(p)}$ and $e^{-i 2 \omega(p)}$
and which includes $-1$ (see Fig.~\ref{f:setsfig}).
With the same procedure we find that
$\Omega^{+-}_f$
is the arc connecting
$e^{i (2 \omega(p+\tfrac{\pi}{2})-\pi )}$ and $e^{-i (2
  \omega(p+\tfrac{\pi}{2})-\pi )}$
which includes $1$ (see Fig.~\ref{f:setsfig}).
We now verify that $\Omega^{++}_f$
and $\Omega^{+-}_f$ are disjoint.
Since $\omega(p) < \tfrac{\pi}{2}$,
$\omega(p+\tfrac{\pi}{2}) > \tfrac{\pi}{2}$
we have
$e^{-i \omega} \in \Omega^{++}_f $
iff
$\omega \mod 2\pi \in (-\pi, -2\omega(p)] \cup [2 \omega(p),\pi] $
and
$e^{-i \omega} \in \Omega^{+-}_f $
iff
$\omega \mod 2\pi \in [ \pi -  2\omega(p + \tfrac{\pi}{2}),  2\omega(p
+ \tfrac{\pi}{2}) -\pi,   ] $.
Then, from the inequality $|\tfrac{\dd}{\dd x} \omega(x)| <1 $, $\forall x \in
\mathbb{R} $, we have
\begin{align*}
\omega(p + \tfrac{\pi}{2}) - \omega(p) =   \int_p^{p+\tfrac{\pi}{2}}
  \! \!\!\! \!\!\!   \!\!\!   \dd {x} \,\,
\frac{\dd}{\dd x} \omega(x)\, <    \int_p^{p+\tfrac{\pi}{2}}
 \! \!\!\! \!\!\!   \!\!\!   \dd {x}  \,\, < \frac{\pi}{2},
\end{align*}
which implies that the sets
$ (-\pi, -2\omega(p)] \cup [2 \omega(p),\pi] $
and
$ [ \pi -  2\omega(p + \tfrac{\pi}{2}),  2\omega(p
+ \tfrac{\pi}{2}) -\pi,   ] $ are disjoint.

Let us now consider the set
$\Omega^{++}_0$.  For $\pi \neq 0, \pi $, the function
$\mathbb{R} \ni k_I \mapsto \omega_{++}(p,i k_I) = \omega(p+ik_I) + \omega(p-ik_I)$
is  smooth.
Therefore, the extremal points of its range occur either in its
stationary points or at its limiting values for
$k_I \to \pm \infty $.
By setting $\partial_{k_I} \omega_{++}(p,i k_I) = 0 $
we obtain
\begin{align*}
  \frac{\sin(p+ik_I)}{\sqrt{1-\nu^2\cos^2(p+ik_I)  }}
  =
  \frac{\sin(p-ik_I)}{\sqrt{1-\nu^2\cos^2(p-ik_I)  }} \implies \\
  \sin^2(p+ik_I) = \sin^2(p-ik_I) \implies \\ \sin(p+ik_I) = \pm \sin(p-ik_I)
  \implies k_I =0,
\end{align*}
where we used the hypothesis $p \neq z \tfrac{\pi}{2}$.
When $k_I = 0$ we clearly have  $\omega_{++}(p, 0) = 2 \omega(p)$.
Let us now compute $\lim_{k_I \to +\infty} \omega_{++}(p, i k_I) $.
Since  $\omega_{++}(p, i k_I) $ is an even function of $k_I$
the limit $k_I \to +\infty$ and $k_I \to - \infty$ coincide.
We have
\begin{align*}
  \omega_{++}(p, i k_I)  = \omega(p+ik_I) + \omega(p-ik_I) =\\
  = 2 \Re \, \omega(p+ik_I) =  2 \Re \, \mathrm{Arccos} (\nu \cos(p+ik_I)) =\\
  =2 \Re \, \mathrm{Arccos} (\nu \cos p  \cosh k_I - i \sin p \cosh k_I) =\\
  =2\, \mathrm{Arccos}  \, 2^{-1}
  \left (
\sqrt{(1+ \cos p  \cosh k_I)^2+ \sin^2 p \cosh^2 k_I} -
  \right. \\
  \left. +\sqrt{(1- \cos p  \cosh k_I)^2+ \sin^2 p \cosh^2 k_I}
  \right ) \xrightarrow{k_I \to +\infty}\\
  \to 2\, \mathrm{Arccos} \cos p = 2 | p |.
\end{align*}
Since we are assuming  $p \in (0, \tfrac{\pi}{2})$
we have that
\begin{align*}
  \frac{\dd}{\dd p } (\omega(p) - p) =
  \frac{\dd}{\dd p }\omega(p) - 1 < 0,\\
  \omega(0) > 0  \mbox{ and }  \omega(\frac{\pi}{2}) = \frac{\pi}{2},
\end{align*}
which imply
$\omega(p) - p  > 0$ for
 $p \in (0, \tfrac{\pi}{2})$.
Similarly one can show
$   \omega(p+\tfrac{\pi}{2}) -\tfrac{\pi}{2} < p $
for $p \in (0, \tfrac{\pi}{2})$.
From
$   \omega(p+\tfrac{\pi}{2}) -\tfrac{\pi}{2} < p  <
\omega(p) $
we have that $e^{-i \omega} \in \Omega^{++}_0 $
iff
$\omega \mod 2\pi \in (-2\omega(p), ), -2 p $.
% which proves that $\Omega^{++}_0$,
% $\Omega^{++}_f$ and $\Omega^{+-}_f$
% are disjoint sets.
Moreover we have that $ e^{ -i \omega} \in \Omega^{++}_2 $ iff
$ e^{ i \omega} \in \Omega^{++}_0 $ iff
$\omega \mod 2\pi \in (2p, 2\omega(p) ) $.  This proves that, for
$p \in (0, \tfrac{\pi}{2})$ $\Omega^{++}_0$, $\Omega^{++}_f$,
$\Omega^{+-}_f$ and $\Omega^{++}_2$ are disjoint sets (see
Fig. \ref{f:setsfig}).  Following the same derivation it is easy to
show that $e^{-i \omega} \in \Omega^{+-}_1 $ iff
$\omega \mod 2\pi \in (-2p, \pi-2\omega(p+\frac{\pi}{2}) ) $ and
$e^{-i \omega} \in \Omega^{+-}_{{-1}} $ iff
$\omega \mod 2\pi \in (2\omega(p+\frac{\pi}{2}) - \pi, 2p ) $ which
proves item $2$ of Lemma~\ref{lem:setsandsol} for
$p \in (0, \tfrac{\pi}{2})$ (see Fig. \ref{f:setsfig}).  The same line
of derivation can be followed for the cases
$p \in (- \tfrac{\pi}{2},0)$, $p \in (\tfrac{\pi}{2} , \pi)$ and
$p \in (-\pi, -\tfrac{\pi}{2})$ thus completing the proof.

\subsection{Proof of item~\ref{i:3}}
Let us consider
a value
$e^{-i \omega} \neq e^{\pm i 2 p}$.
From item $2$ of Lemma \ref{lem:setsandsol} we have that the sets
 $\mathrm{\Omega}^{++}_{f} ,
    \mathrm{\Omega}^{+-}_{f} ,
    \mathrm{\Omega}^{++}_{0} ,
    \mathrm{\Omega}^{++}_{2} ,
    \mathrm{\Omega}^{+-}_{-1} ,
    \mathrm{\Omega}^{+-}_{1} $
    cover the whole unit circle except the points
    $e^{\pm i 2 p}$ and therefore
    $e^{-i \omega}$ must belong to one of those sets.
 We prove the thesis for the case
     $e^{-i \omega} \in \mathrm{\Omega}^{++}_{f} $, and the remaining cases
     can be proved in the same way.
     If  $e^{-i \omega} \in \mathrm{\Omega}^{++}_{f} $,
     then there exists $k \in \Gamma_f$ such that
$\omega_{++}(p,k) = \omega \mod 2\pi$.
By direct computation one verify that also $\omega_{++}(p,-k)=
\omega_{--}(p, k -\pi ) = \omega_{--}(p, \pi - k ) $ $\mod 2 \pi = \omega$
In order to prove that these are the only admissible solutions
we must check that $k'\neq \pm k $ implies
$\omega_{++}(p,k') \neq \omega \mod 2\pi$
By contradiction let us suppose that there exists  $k'\neq \pm k $
such that
$\omega_{++}(p,k') = \omega \mod 2\pi$. This clearly implies
$\omega_{++}(p,k') = \omega_{++}(p,k')$ since the range of
$\omega_{++}$ is smaller than $2 \pi$.
Let us consider the case $ 0< k' < k$.
Since $\omega_{++}$ is smooth, there must exists
$k''$ such that $ k'< k'' < k$
and $[\tfrac{\dd}{\dd k } \omega_{++}] (p,k') =0$.
By direct computation one proves that this is impossible.
The generalization to the cases
$ -k < k' <0$,
$   k < k' < \pi $,
$     -\pi<k'<k $ is straightforward. The analysis of the
cases
$ k' = 0,\pi $ is easily done by direct computation.

\section{Proof of Lemma~\ref{lem:boundstates}}\label{a:lemma-boundstates}
In order to prove the lemma it is convenient to introduce the
following function from the negative half line $k_I \in (-\infty,0]$
to the unit circle $S^1$:
\begin{equation*}
\begin{aligned}
&G_{z}: \mathbb{R}^{-}\rightarrow S^1,\quad
  G_z(k_I)= -\frac{A_z(p,k)}{A_z^*(p,k)},\qquad
j=0,2,\pm 1,\\
&A_0(p,k)=\sin(\omega(p-ik_I)) +  \nu \sin(p-ik_I),\\
&A_2(p,k)=\sin(\omega(p-ik_I)) -  \nu \sin(p-ik_I),\\
&A_{\pm 1} (p,k)=\sin(\omega(p\mp \frac{\pi}{2}- ik_I)) +  \nu \sin(p\mp\frac{\pi}{2}- ik_I).
\end{aligned}
\end{equation*}
The above function allows to study the constraint $T_\pm(p,k,\chi)=0$
through the following five properties
\begin{enumerate}[(a)]
\Item \label{it:1}
\begin{equation*}
\begin{aligned}
&k\in\Gamma_0: T_+(p,k,\chi)=0\iff G_0(k_I) = e^{i\chi}, \\
&k\in\Gamma_2: T_+(p,k,\chi)=0\iff G_2(k_I) = e^{i\chi}, \\
&k\in\Gamma_{\pm1}: T_-(p,k,\chi)=0\iff G_{\pm 1}(k_I) = e^{i\chi}, \\
\end{aligned}
\end{equation*}
\Item \label{it:2}
\begin{equation*}
\begin{aligned}
G_{0}(0) =G_{2}(0)= -1,\qquad G_{\pm 1}(0)= 1,
\end{aligned}
\end{equation*}
\Item \label{it:3}
\begin{equation*}
\forall k_I,\:\: G_{0}(k_I)\neq 1,G_{2}(k_I)\neq 1,\:\:
                        G_{\pm 1}(k_I)\neq -1,
\end{equation*}
\Item   \label{it:4}
\begin{equation*}
\begin{aligned}
 \forall k_I, \quad \frac{d}{d k_I} G_z(k_I)\neq 0,\quad
z=0,2,\pm 1,
\end{aligned}
\end{equation*}
\Item \label{it:5}
\begin{equation*}
\begin{aligned}
\lim_{k_I \to - \infty} G_0(k_I) = e^{i 2|p|},\\
\lim_{k_I \to - \infty} G_2(k_I) = e^{-i 2|p|},\\
\lim_{k_I \to - \infty} G_{\pm 1}(k_I) =
  \begin{cases}
e^{\pm i2p},
&
p \in (-\frac{\pi}{2},\frac{\pi}{2})\\
e^{\mp i2p},
&
p \in (- \pi,-\frac{\pi}{2}) \cup (\frac{\pi}{2},\pi)
  \end{cases}.
\end{aligned}
\end{equation*}
\end{enumerate}
The statement of the lemma is now proved noticing that the functions
$G_z(k_I) $, $z=0,2,\pm 1$, from the negative half line
$k_I \in (-\infty,0]$ to the unit circle $S^1$, are injective and
their ranges are
\begin{itemize}
\item
Range of $G_{0}$:  the smallest arc having $-1$ and $ e^{i 2 |p|}$
    as its extremal points,\\
\item
Range of $G_{2}$:  the smallest arc
    having $-1$ and $ e^{-i 2 |p|}$ as its extremal points,\\
  \item Range of $G_{\pm 1}$: the smallest arc having $1$ and
    (depending on the value of $p$) $e^{ \pm i 2 |p|}$ as its extremal
    points.
\end{itemize}

Here we provide the proofs of items 1-5 for the case
$k \in \Gamma_0$. The proof for the other three cases $k \in \Gamma_2$
and $k \in \Gamma_{\pm 1}$ is almost identical.

\subsubsection{Proof of item~\eqref{it:1} for  $k \in \Gamma_0$}
If $k \in \Gamma_0$ it is $k=ik_I$, and starting from the definition of
Eq.~\eqref{eq:solclass2} we can rewrite $T_+(p,k,\chi)$ as follows
\begin{equation*}
\begin{aligned}
 T_+ = \frac{  A^*_0(p,k)+ e^{-i \chi} A_0(p,k)}
{ e^{-i \chi} A^*_0(p,k)+ A_0(p,k)}.
\end{aligned}
\end{equation*}

Let us replace $A_0(p,k)$ with $A$ and $T_+(p,k,\chi)$ with $T_+$ in
order to lighten the notation. We have that
$T_+=0\iff A^*+ e^{-i \chi} A=0 \land e^{-i \chi} A^*+ A\neq 0$.
First we observe that $A \neq 0 $, indeed
$A=0 \iff \sin(\omega(p-ik))=-n \sin(p-ik) \implies
\sin^2(\omega(p-ik))=n^2 \sin^2(p-ik) \implies \nu^2 =1$,
which is not an admissible value. Accordingly, a straightforward
computation shows that
$T_+=0\iff A^*+ e^{-i \chi} A=0 \land e^{-i \chi} A^*+ A= 0$ iff
$\chi=m\pi$ ($m\in\mathbb{Z}$). However, $T_+(p,k,m\pi)=(-1)^m\neq 0$
and we conclude that $T_+=0$ if and only if $A^*+ e^{-i \chi} A=0$
which proves item~\eqref{it:1}.

\subsubsection{Proof of item~\eqref{it:2} for  $k \in \Gamma_0$}

Notice that $G_0(k_I)= -1 \implies A = A^*$ which implies
$ \Im[A]=0 \implies \Im[A^2 ]=0$ (where we replaced $A_0(p,k)$ with
$A$ in order to lighten the notation).  Since we have
$A^2 =1-\nu^2 +2\nu\sin(p-ik_I)A$ the condition
$\Im[A]=0 \land \Im[A^2 ]=0$ implies $\Im\sin(p-ik_I)=0$ that is
$\cos(p)\sinh(k_I)=0$ and then $k_I=0\, \lor\, p=\frac{\pi}{2}+m\pi$
($m\in\mathbb{Z}$) which proves item~\eqref{it:2}.

\subsubsection{Proof of item~\eqref{it:3} for  $k \in \Gamma_0$}

We have $G_0(k_I)= 1 \implies A = -A^*$ which implies
$ \Re[A]=0 \implies \Im[A^2 ]=0$ (where we replaced $A_0(p,k)$ with
$A$ in order to lighten the notation).  Since it is
$A^2 =1-\nu^2 +2\nu\sin(p-ik_I)A$ the condition
$\Re[A]=0 \land \Im[A^2 ]=0$ implies $\Re\sin(p-ik_I)=0$ that is
$\sin(p)\cosh(k_I)=0$. Since the last equality is satisfied only for
$p=m\pi$ ($m\in\mathbb{Z}$), which are not admissible values of $p$,
item~\eqref{it:3} is proved.

\subsubsection{Proof of item~\eqref{it:4} for  $k \in \Gamma_0$}
We prove that $\frac{d}{dk_I}G_0(k_I)= 0 \implies p = m \frac{\pi}{2}$
$(m \in \mathbb{Z} )$, which are not admissible values of $p$. Again
we replace $A_0(p,k)$ with $A$ in order to lighten the notation.
Consider
\begin{align}\nonumber
  \frac{d}{dk_I}G_0(k_I) = \frac{A'A^*-A{A'}^*}{(A^*)^2}
\end{align}
where $A'=\partial_{k_I}A$ and ${A^*}':=\partial_{k_I} A^*={A'}^*$.
Then, reminding that $A\neq 0$ (see the proof of item~\eqref{it:1})
and noticing that
\begin{equation*}
\begin{aligned}
& A'A^*-A{A'}^*=-i |1+ \omega'_-|^2\sin(\omega_-+\omega_+),\\
& \omega_{\pm}:=\omega(p \pm ik),\quad \omega'(x) := \frac{d}{dx}\omega(x),
\end{aligned}
\end{equation*}
(this can be verified rewriting $A$ as
$A= \sin(\omega(p-ik))(1+\omega'(p-ik))$) one has
\begin{equation}\nonumber
  \frac{d}{dk_I}G_0(k_I)=0 \iff 1+ \omega'_- =0 \lor \sin(\omega_-+\omega_+)=0.
\end{equation}
Let us investigate the two possible cases.  In the first case
$1+ \omega'_- =0$ it must be
\begin{align*}
   {\omega'_- }^2 = \frac{\nu^2
  \sin^2(p-ik_I)}{\sin^2(\omega(p-ik_I))}=1\implies \nu=1,
\end{align*}
which is not admissible. Let us now consider the case
$\sin(\omega_-+\omega_+)=0$, that is $\omega_-+\omega_+ = m \pi$.  We
have for $m$ even
$
\cos \omega_- = \cos\omega_+
 \implies\sin p=0 \lor k_I = 0.
$
On the other hand, if $m$ is odd we have
$
\cos \omega_- = -\cos\omega_+\implies
\cos p=0 .
$
Item~\eqref{it:4} is thus proved.

\subsubsection{Proof of item~\eqref{it:5} for  $k \in \Gamma_0$}

For convenience in the following we replace $A_0(p,k)$ with $A$.
First we rewrite the function $G_0$ as follows
\begin{align*}
& G_0(k_I) = -\frac{Z}{Z^*},\\
&  Z := -iA  = e^{-i\omega(p-ik_I)}-\nu e^{i(p-ik_I)}.
\end{align*}

Reminding that in this case it is $k\in\Gamma_0$ it is $k=k_R+i k_I$
with $k_R=0$, from Appendix~\ref{a:eigenvalues} we have the
expressions of $e^{-i\omega(p-ik_I)}$ for $k_I \to -\infty$:
\begin{align*}
  p>0& \implies e^{-i\omega(p-ik_I)}=\frac{1}{\nu}e^{-ip}e^{k_I},\\
  p<0 &\implies e^{-i\omega(p-ik_I)}=\nu e^{ip}e^{-k_I}-\frac{\mu^2}{\nu}e^{-ip}e^{k_I}.
\end{align*}
from which item~\eqref{it:5} follows.

\section{Asymptotic behaviour of the walk eigenvalues}\label{a:eigenvalues}
The one-particle Dirac walk in momentum space is defined through the
matrix valued function of Eq.~\eqref{eq:free-qw}. Since
$U(p) \in \mathbb{SU}(2)$, its eigenvalues are $e^{-i \omega(p)}$ and
$e^{i \omega(p)}$ where $\omega(p)$ is the solution of the equation
$\cos\omega =\nu \cos p$ with positive value.  Then we write
\begin{align}\nonumber
  \omega &: (-\pi,\pi] \to [0,\pi]
           & p \mapsto \omega(p) = \arccos(\nu \cos p)\geq 0.
\end{align}
For our purposes it is convenient to consider the analytic
continuation of $U(p)$ to the subset
$\mathcal{S} := \{p\in \mathbb{C} \,|\, \Re(p)=p_R \in
(-\pi,\pi],\Im(p)=p_I \leq 0 \}$
of the complex plane.  The eigenvalues of $U(p)$, with
$p \in \mathcal{S}$, are $e^{-i \omega(p)}$ and $e^{i \omega(p)}$
where now $\omega(p) =\mathrm{Arccos}(\nu \cos p)$, and
$\mathrm{Arccos}$ denotes the principal value of the multivalued
analytic function $\arccos$.  We notice that
$\mathrm{Arg}(e^{i \omega(p)}) = \Re(\omega(p)) =
\Re(\mathrm{Arccos}(\nu \cos p) \in [0,\pi] $.

In the two-particles case we introduced the center of mass coordinates
$p,k$, representing respectively the total and the relative
momentum. While $p$ is always real, $k$ can have an imaginary
part. Let us study the eigenvalues of $U(p-k)$ in the limit
$k_I \to -\infty $. We have
\begin{align}\nonumber
 & U(p-k)=\\
 & \nu e^{i(p-k_R)}e^{-k_I}
  \begin{pmatrix}
    e^{2k_I} & -i\frac{\mu}{\nu}e^{-i(p-k_R)}e^{k_I} \\\nonumber
    -i\frac{\mu}{\nu}e^{-i(p-k_R)}e^{k_I} & e^{-i2(p-k_R)}
  \end{pmatrix},
\end{align}
and denoting with $\lambda'_1$,$\lambda'_2$ the two eigenvalues of
$\nu^{-1}e^{-i(p-k_R)}e^{k_I} D(p-k)$ we have for
$k_I \to -\infty $
\begin{align}\nonumber
  \lambda'_1= 1-\frac{\mu^2}{\nu^2}e^{-2i(p-k_R)}e^{2k_I},\quad
  \lambda'_2= \frac{1}{\nu^2}e^{-2i(p-k_R)}e^{2k_I}.
\end{align}
Accordingly, for $k_I \to -\infty $, the eigenvalues $\lambda_1$ and  $\lambda_2$
of   $ D(p-k)$ are
\begin{align}\nonumber
  &\lambda_1= \nu e^{i(p-k_R)}e^{-k_I}-\frac{\mu^2}{\nu}e^{-i(p-k_R)}e^{ k_I},\\\nonumber
  &\lambda_2= \frac{1}{\nu}e^{-i(p-k_R)}e^{ k_I},
\end{align}
and noticing that
$ \lim_{k_I \to -\infty } \mathrm{Arg}(\lambda_1) = p-k_R$, and
$\lim_{k_I \to -\infty } \mathrm{Arg}(\lambda_2) = -(p-k_R)$ we
get
\begin{equation}\nonumber
 \begin{aligned}
&   p-k_R >0 \implies e^{-i \omega (p-k)} = \lambda_2,\: e^{i \omega (p-k)} = \lambda_1,\\
&   p-k_R <0 \implies e^{-i \omega (p-k)} = \lambda_1,\: e^{i \omega (p-k)} = \lambda_2.
 \end{aligned}
\end{equation}

\end{document}